# Validation Workflow for Machine Learning Interatomic Potentials for Complex Ceramics

Kimia Ghaffari[1], Salil Bavdekar[2], Douglas E. Spearot[1], Ghatu Subhash[1,*]

[1] Department of Mechanical and Aerospace Engineering, University of Florida, Gainesville, FL 32611 USA

[2] Department of Materials Science and Engineering, University of Florida, Gainesville, FL 32611 USA

## Abstract

The number of published Machine Learning Interatomic Potentials (MLIPs) has increased significantly in recent years. These new data-driven potential energy approximations often lack the physics-based foundations that inform many traditionally developed interatomic potentials and hence require robust validation methods for their applicability, accuracy, computational efficiency, and transferability to the intended applications. This work presents a sequential, three-stage workflow for MLIP validation: (i) preliminary validation, (ii) static property prediction, and (iii) dynamic property prediction. This material-agnostic procedure is demonstrated in a tutorial approach for the development of a robust MLIP for boron carbide ($B_4C$), a widely employed, structurally complex ceramic that undergoes a deleterious deformation mechanism called 'amorphization' under high-pressure loading. It is shown that the resulting $B_4C$ MLIP offers a more accurate prediction of properties compared to the available empirical potential.

*Corresponding author: G. Subhash (subhash@ufl.edu)

# 1. Introduction

Atomistic simulations have unlocked access to nano-level observation and prediction of material behavior. These studies lack many of the physical and fiscal restrictions of their experimental counterparts and hence enable the study of materials before they are synthesized in a laboratory and the analysis of materials subjected to complex boundary conditions. Evaluation of material viability for extreme applications (ballistic, nuclear, aerospace, etc.) is heavily reliant on computer simulations as the necessary temperatures and pressures are difficult to achieve, if not impossible to replicate experimentally. Atomistic simulations such as molecular dynamics (MD) rely on interatomic potentials (IPs), which describe the potential energy surface (PES) of a material and hence can be used to compute interatomic forces under any deformation conditions. An ideal IP must satisfy three main requirements: (i) accuracy, (ii) transferability, and (iii) computational efficiency during runtime. The accuracy of an IP is directly responsible for the ability of a simulation to capture underlying physics, its transferability enables the investigation of more dynamic environments (beyond the trained environment and data) where diverse atomic configurations may be encountered, and its computational efficiency allows for larger and longer simulations at increased speed for observation of behaviors beyond the nanoscale. Unfortunately, traditional IP development methods are often limited in their ability to effectively satisfy the above three requirements.

*Ab initio* or quantum mechanics (QM) based methods, such as Density Functional Theory (DFT), model the PES by solving for the electronic structure of a material based on atomic species and their relative positions using an approximation of Schrödinger's equation. This method can be both accurate and transferable, as it relies on laws of physics to make predictions of material properties and behaviors. Unfortunately, this method comes with a very high computational cost, with the largest systems being limited to 1000s of atoms for ps time durations [1]. Empirical or semi-empirical models (Lennard-Jones [2], Embedded-Atom Method [3], Stillinger-Weber [4], etc.) are constructed with a relatively simple functional form and parameters fitted so that the simulations reproduce material properties (bonding behavior, vibrational properties, thermodynamic behavior, etc.). These models greatly simplify energy and force calculations by approximating the PES of a material system, allowing for simulations of millions of atoms for longer time durations. However, this approach comes at the cost of transferability, with a single empirical model being developed for use in a narrow range of simulation conditions. Moreover, the development of empirical models is non-trivial because it often needs extensive knowledge of materials science and chemistry for parameter selection and fitting. This necessary human intervention in classical IP development coupled with their sometimes-limited transferability results in a higher barrier-to-entry for the computational evaluation of promising, novel materials for extreme applications. Until recently, *ab initio* and empirical models were the two dominant methods of describing the PES of a material system, but as the need for modeling more complex materials emerges, the demand for more flexible IPs increases.

Due to rapid improvements in computing power and database availability, machine learning (ML) methods have exploded in popularity across nearly every scientific field of study. Computational materials modeling is no exception, with machine learning interatomic potentials (MLIPs) promising to bring almost *ab initio* accuracy at the speed of empirical potentials [5]. MLIPs can extract the intrinsic relationship between atomic configuration and potential energy using statistical learning algorithms. Such statistical algorithms can be transferable across a wide range of material systems and deformation conditions at an increased computational efficiency.

This approach to IP development is particularly useful when applied to the modeling of ceramics with complex crystal structure. The diverse energy landscape of such structurally complex ceramics can challenge the already limited transferability of traditional empirical potentials. However, MLIPs can leverage their inherent flexibility to learn implicit energy-configuration relationships efficiently and deliver a robust model for prediction of material response.

The data-driven nature of ML methods naturally breeds doubt in their capability to capture physics and not just memorize trends in the training database. MLIPs have shown remarkable success in accurately and efficiently approximating the PES of Si [6], Ti [7], GeTe (phase change materials) [8], and the Si-C-N system [9]. However, in many published articles, minimal details are provided on how the potential is developed, what kind of training data is needed, the influence of various parameters on the MLIP development process, and its transferability to more complex situations. Many decisions are critically important at various stages of MLIP development (e.g., selection of training data composition, data representation, simulation domain sampling) and due to the "black-box" nature of some ML-methods the precise model creation pathway should be explicitly described to elucidate the effects of such variables on the final potential. Accordingly, the goal of this work is to shed more light on the details of the development process, the breadth of training database diversity needed, and its predictability beyond the trained environment, with a specific focus on complex, non-unary ceramic systems. The objectives of this work are twofold: (i) Define a robust, material-agnostic procedure for development and validation of a MLIP for structural ceramics with complex crystal structure and (ii) apply this process to a well-studied ceramic system to provide a robust generalized framework for expansion to other complex systems.

In particular, our interest is in boron-icosahedral ceramics, which have a unique crystal structure with a 12-atom icosahedron and a three- or two-atom chain in a unit cell. In addition, icosahedral ceramics also exhibit polymorphism. For example, boron carbide has 52 polymorphic structures [10] and a synthesized material may contain many of these polymorphs at different volume fractions. These ceramics have a desirable combination of properties which allows them to be used in various extreme conditions [11-15]. For example, the open-caged structure and highly covalent bonding provide low density and high hardness [16] which are ideal for impact applications [13], while its high neutron absorption and high-temperature stability is promising in nuclear radiation shielding applications [17]. Computational simulations of boron-icosahedral ceramics under extreme environments, such as shock loading, are necessary to test their viability without resource-intensive experiments. Unfortunately, the barrier for atomistic analysis of these advanced ceramics remains high due to their complex structure, energy description, and deformation response. Among the boron-icosahedral ceramics, only boron carbide ($B_4C$) has an established IP [18]; this ReaxFF potential has been used to predict material behavior in shock environments [19-21]. However, concerns about the applicability of ReaxFF potentials across a broad range of atomic environments (bulk, surface, cluster, etc.) have been raised [22]. Recently, An et al. have demonstrated some of the promising capabilities of icosahedral-ceramic NN-based MLIPs under various conditions in large-scale simulations [23-25]. Unfortunately, the computational cost associated with the generation of large training databases ($N_{samples}$~ 270,000 to 1,200,000 [26-29]) often used in training complex NN interatomic potentials continues to limit their rapid development. Thus, there is a need for a systematic IP development method to minimize superfluous database additions for $B_4C$ and other complex ceramics while maintaining sufficient accuracy and transferability. Through our
5

proposed validation approach, we train a highly accurate NN-based $B_4C$ MLIP with a significantly reduced training database size of ~39,000 samples.

Ultimately, this manuscript takes a tutorial-like approach to provide a robust and material-agnostic MLIP development process for complex ceramics and is organized as follows. Section 2 outlines and reviews MLIP ingredients and provides a development workflow that is then applied to the boron carbide system. Section 3 presents the results of the MLIP workflow for $B_4C$, followed by discussion on the data requirement, predictive power, and transferability in Section 4. Section 5 summarizes the findings of the manuscript and provides major conclusions.

## 2. Method Development

In general, the MLIP development process consists of four major components: (i) training database, (ii) regression model, (iii) data representation, and (iv) model validation. This section discusses the importance of each component, provides best practices, and applies them to $B_4C$ MLIP development.

### 2.1 Training Database

The quality of any ML model is directly linked to the quality of its training database. In the case of MLIP development, a high-quality training database must contain atomic configurations that are reasonably representative of all anticipated configurations in the intended simulation environments. Thus, the data set for MLIP training consists of atomic species and coordinates, their associated total energy, forces, and virial stresses for a broad range of atomic configurations (e.g., various strain levels in tension, compression, and shear loading as well as atomic configurations in different material phases). This data is generated with ab initio methods to maximize accuracy, though supplementation can be made with data generated with empirical potentials for computational efficiency (assuming a sufficiently accurate empirical IP is available). The specific structures in the database should be curated to have sufficient energetic diversity within the intended simulation domain. This allows the model to see different atomic configurations, thus increasing the transferability of the MLIP under unknown or more complex loading scenarios. It should be noted that higher energetic diversity in the training database can require much larger databases for adequate learning, which is computationally costly to generate. Thus, there is a balance between the inclusion of sufficient energetic diversity and the minimization of the training database size depending on the intended final application.

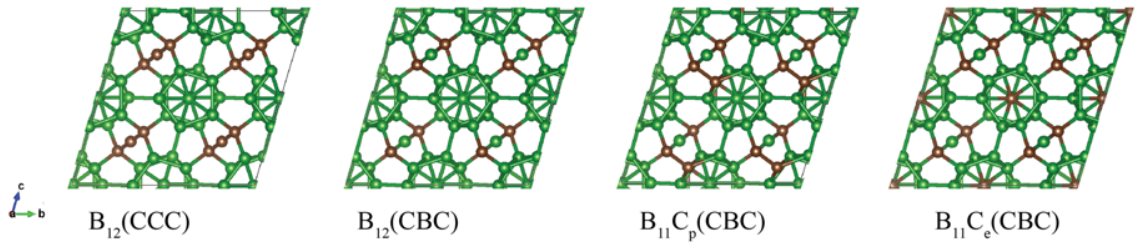

Figure 1. The four most prevalent and energetically favorable $B_4C$ polytypes used in MLIP training data. Boron atoms are shown in green with C atoms shown in red. The notation in the ( ) refers to the chain between boron icosahedra.

The behavior of $B_4C$ in extreme environments, especially during high-rate and shock-loading applications, is the focus of our study, thus the MLIP training database is curated for those conditions. Due to the size similarity between boron and carbon, they both can occupy either chain or icosahedral positions. This results in 52 energetically stable $B_4C$ polytypes, and among these the four most prevalent structures being $B_{12}(CCC)$, $B_{11}C_e(CBC)$, $B_{11}C_p(CBC)$, and $B_{12}(CBC)$ [10], each of which are shown in Figure 1. Thus, we include these four polytypes to capture a range of possible environments. 2 x 2 x 2 supercells (120 atoms) of these structures are equilibrated in an NVT ensemble using *ab initio* MD (AIMD) calculations at various temperatures (5 K to 3650 K in steps of 675 K) for 1 ps, with snapshots taken every 1 fs. Snapshots of various modes of deformation (shear, uniaxial, and volumetric) are included for each polytype to increase the energetic diversity in the training set. All data is generated using the projector-augmented-wave (PAW) method [30, 31] with the Vienna Ab Initio Simulation Package (VASP) [32-35] software. The exchange-correlation energy is modeled by the Perdew-Burke-Ernzerhof (PBE) functional [36]. A cutoff energy of 700 eV is selected for the plane-wave basis set, using the tetrahedron method with Blöchl corrections with 8 k-points for the Brillouin zone integration. The structural relaxations are carried out with tolerances of $10^{-6}$ eV for electronic convergence and $10^{-5}$ eV for ionic convergence. The resulting database consists of 39,083 total snapshots that are randomly shuffled before being divided into a 90%-10% training-testing split.

## 2.2 Choice of regressor

The core of ML-based methods is the learning algorithm, or regressor, used to map the input to the output. There are several types of regressors, each with their respective advantages and challenges. For MLIP development, most force fields fall in one of three groups: (i) linear regression models, (ii) kernel-based models, and (iii) neural network (NN)-based models. Linear regressor force fields (e.g., SNAP [37], MTP [38], UF$^3$ [39]) are based on a linear combination of input features of the system of interest. By leveraging the bulk of computational resources up-front on descriptor formulations, such models can offer extremely computationally efficient force fields with training times on the order of seconds [39]. A caveat to this efficiency is the reliance on feature extraction of the system, with more complex systems (i.e., those with many energy contributions) often requiring large feature sets and can thus be prone to overfitting. Kernel-based MLIPs (e.g., GAP [40], AGNI [41]) use similarity functions, or kernels, to predict atomic forces and energies based on structural environments in the reference database. These models are ideal for systems with smaller training databases, however choosing and optimizing kernels can require significant human intervention which increases development time and may reduce transferability. Kernel-based models can also become memory-intensive as they store reference configurations for use in energy predictions during run-time.

On the other hand, NN-based models mimic the biological learning processes of the brain and consist of a series of interconnected layers of nodes (neurons) with each node having an activation value determined by a weighted sum of nodes from the previous layer. The weights associated with node-node connections are optimized via a backpropagation learning algorithm [42] during model training. This node-layer architecture allows the NN to "activate" or "inhibit" specific neural pathways to learn complex relationships not necessarily discernible by human observation. The flexibility of NN architectures also allows for learning of complex functional forms of the PES and other parameters. NNs with several hidden layers are "deep" learners, and thus do not require feature engineering to the extent that "shallow" methods like kernel-based models do. These benefits come with a cost, as most NNs require substantially larger training data sets and are "black-box" in nature, lacking the physical interpretability of their shallow-learning counterparts. Fortunately, with growing computing power and open-access databases, access to large amounts of training data for MLIPs has become less prohibitive.

Many MLIPs for complex materials rely on the flexibility and hands-off nature of NNs to capture hidden features within atomic structures to predict energies in a variety of environments. For example, Huang et al. [43] developed a deep learning potential for boron subphosphide ($B_{12}P_2$) capable of capturing nanotwinning and other defects that have been widely observed in similar boron-icosahedral materials [16, 44]. We thus employ a deep, densely connected, NN model (via DeePMD-kit [45]) for learning the $B_4C$ PES to maximize flexibility for a dynamic environment under extreme conditions and minimize human intervention that would be necessary for future implementation in other boron icosahedral ceramics subjected to complex loading.

Despite its flexibility, the same NN model may not be ideal for all systems, and hyperparameters must be chosen to optimize the regressor for the given system. The model architecture (activation function, number of layers and neurons), choice of weight regularization, learning rate, and loss-function coefficients are particularly influential. There are several methods available for automatic hyperparameter optimization in the literature [46-48]. To demonstrate the tuning process, results for a manual optimization of these parameters for the $B_4C$ system are given and discussed in Section 3.1.

## 2.3 Data Representation

Data representation describes the local atomic configuration in a machine-readable format (a descriptor) and captures the relevant features for the given system (3-body terms, polarity, long-range interactions, etc.). What features are considered relevant will depend on the regressor model, as more flexible models (like deep NNs) can intrinsically extract important features. An ideal descriptor of atomic configuration is invariant with respect to translation, rotation, and atomic permutation, while also being smooth (differentiable) and unique [49, 50]. Several off-the-shelf packages are available for processing of raw, ab initio-generated data to machine readable descriptors [51]. Various descriptors are available (e.g., Coulomb Matrix [52], SOAP [53], ACSF [54]) with the biggest differentiator being their resolution; global descriptors like the Coulomb Matrix encode the entirety of the system, while local descriptors like SOAP and ACSF represent local atomic configuration based on provided cutoff radii (the radius within which an atom influences its neighbors) and/or neighbor lists.

The DeePMD-kit descriptor uses symmetry functions (similar to ACSF) and an embedded neural network to map atomic positions to a descriptor. The high resolution of this descriptor is ideal for the observation of highly localized, dynamic effects present under extreme conditions, while the inclusion of an embedding network allows the descriptor mapping to learn and evolve during training. This descriptor also consists of several tunable hyperparameters: embedding network architecture, activation function, choice of interaction terms (2-body, 3-body, polarity), cutoff radii, axis neuron, and a maximum number of neighbors. Specifically, for $B_4C$, we choose to include all information (radial and angular) for a 2-body embedding descriptor to capture the directionality of atomic bonding in the crystal structure. Other hyperparameters for the descriptor are chosen based on tuning experiments (Section 3).

## 2.4 Model Validation

Validation of the MLIP is crucial for the assessment of its applicability to the desired problem and observation of the extent of transferability beyond the trained environment. In the case of PES-fitted models, their ability to capture various physical and mechanical properties, and behaviors beyond the given reference (training) dataset is required to evaluate the generalizability or transferability of the IP to other environments or modes of deformation. Validation metrics for MLIPs vary in complexity, ranging from general statistical error evaluation (in forces, energies, and virial stresses) to prediction of material properties (lattice constants, elastic constants, thermodynamic properties, etc.).

The precise validation metrics used for each MLIP are dependent on the application of the model, with models for diverse simulation conditions requiring more thorough validation. To provide structure and guidance, we propose a sequential model validation workflow, as depicted in Fig. 2, consisting of three stages with increasing complexity: (i) preliminary evaluation, (ii) static property prediction, and (iii) dynamic property (or behavior) prediction. The sequential nature of this workflow can reduce MLIP development time and increase its accuracy as well as transferability as models that do not pass early validation stages would not progress to later, more computationally challenging, and expensive stages. Specific metrics chosen for each stage (particularly in stage 2 and 3) are left to the user to customize depending on their application. For example, if lattice defects are of interest, they would be included in the training database as well as structures used for the validation workflow. In this case, the formation energies of defects can be included in stage 2 validation, while diffusion kinetics would be an applicable stage 3 metric. This procedure can be automated and integrated into existing on-the-fly validation algorithms [55] to provide a more physics-informed aspect to model vetting.

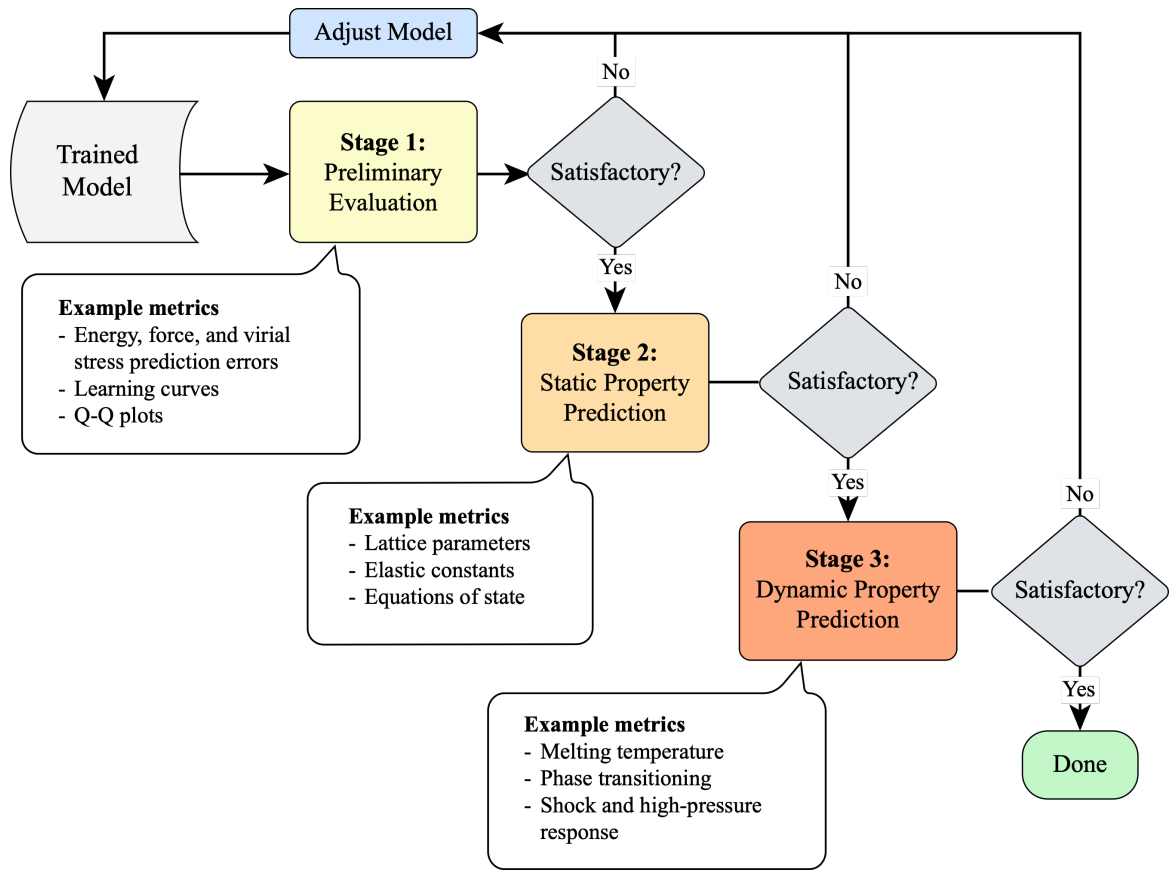

Figure 2. Proposed MLIP validation workflow, along with example metrics, to capture material behavior under shock loading. The model validation metrics at each stage may be different depending on the intended final application of the MLIP. Adjustments can be made to the model after each validation stage if performance is unsatisfactory.

Preliminary evaluation (stage 1) of MLIPs consists of basic analysis of the developed model. Only models that perform well in cross-validated energy, force, and virial stress prediction error and have stable learning curves will move on to subsequent stages where more advanced (and computationally costly) testing can take place. In stage 2, MLIPs are tested for their ability to accurately predict static material properties (i.e., elastic constants, lattice constants, material density, etc.) using atomistic simulation software, such as the Large-scale Atomic/Molecular Massively Parallel Simulator (LAMMPS [56]). The properties calculated using the MLIP can then be directly compared to *ab initio* results and/or experimental results (e.g., elastic constants or wave velocities in various orientations of the crystal). This portion of the testing process will assess the ability of the MLIP to predict properties that it has not explicitly been trained on, confirming that the model has learned the basic physics of bonding for the material system of interest. Finally, dynamic behaviors are computed (stage 3), specifically involving the behavior under various thermal conditions and the making and breaking of bonds. The ability of the MLIP to accurately predict phase changes, thermal transport properties, shock response or other deformation environments, etc., provides adequate final validation that the developed MLIP is ready for use in MD simulations for the intended application (e.g., ballistic impact). These proposed stage 3 measures may involve complex chemical reaction processes and thus are amongst the most robust metrics of a MLIP's predictive power beyond the trained environment.

## 3. B$_4$C MLIP Development

In this section we discuss the specifics of each validation stage discussed in Fig. 2 for a B$_4$C MLIP intended for extreme conditions, and how one set of results inform subsequent adjustments to the model.

### 3.1 Stage 1: Preliminary Evaluation

Preliminary evaluation consists of model hyperparameter optimization using target prediction-based error metrics, and selection of the best performing parameter set (based on root mean square error (RMSE)) for further model development in terms of static and dynamic material property and performance metrics. All initial hyperparameters (Model 1.0 in Table 1) are chosen based on the recommendations given by DeePMD-kit documentation [45]. For a more detailed list of all hyperparameters used, the input scripts of all mentioned models have been made available on GitHub[1].

Automatic hyperparameter optimizations through grid-search methods are ideal for covering a large portion of parameter-space with minimal human intervention. However, this process can become extremely costly for complex regressors (i.e., NNs) with larger parameter sets and longer training times. Thus, we demonstrate the process of manual optimization for three DeePMD-kit NN parameters: (i) embedding network, (ii) cutoff radii, and (iii) fitting network. Table 1 illustrates several MLIP models with varying hyperparameter sets, organized into three sections based on the tuning of these parameters. The embedding network is an additional NN used in generating the descriptor for a series of relative radial distances. The architecture (number of neurons and layers) of this network is analogous to the complexity of the descriptor, with large networks resulting in larger, more complex descriptors and higher computational cost in model training and application. The cutoff radii consist of smooth and hard radial limits for interatomic interactions. Finally, the fitting network is the architecture for the NN mapping the descriptor to its respective energy value, force vector, and virial stress tensor. Further details on all hyperparameters for DeePMD-kit package can be found in [45]. All models share the same training database described in Section 2.1, with shear, uniaxial, and volumetric strain states ranging from -1% to 1%. Each hyperparameter is varied in isolation, with the best performing value persisting in subsequent tunings, as indicated by the bold font in Table 1. The final hyperparameter set of Model 1.10 (Embed: 25-50-100, Fit: 120-120-120, Radii: [0.5, 6.0]) has a significantly lower validation set RMSE than its counterparts compared to DFT values, and thus all subsequent models will use these hyperparameter values. However, it should be emphasized that low numerical errors in MLIP predictions on the validation set do not guarantee good model performance in capturing the physical (and chemical, if relevant) behaviors of a material system in later computations. Thus, additional testing and adjustment of MLIP models are often necessary to ensure generalizability, and these will be discussed in subsequent sections.

---
[1] https://github.com/SubhashUFlorida/B4C-MLIP.git

Table 1. MLIP parameter sets for manual assessment of embedding net architecture, cutoff radii tuning, fitting net, and the resulting Stage-1 values. The bold font numbers indicate the best performing values and are carried forward to the next tuning.

| Model # | Embedding network architecture | Cutoff Radii [$R_{smooth}$, $R_{cut}$] | Fitting network architecture | Energy RMSE (eV/atom) | Force RMSE (eV/Å) | Virial RMSE (eV/atom) |
|---|---|---|---|---|---|---|
| Embedding net | | | | | | |
| 1.0 | 10-20-40 | [4.5, 6.0] | 150-150-150 | 0.0257 | 0.0697 | 0.0329 |
| 1.1 | 25-25-25 | [4.5, 6.0] | 150-150-150 | 0.0209 | 0.0708 | 0.0363 |
| 1.2 | **25-50-100** | [4.5, 6.0] | 150-150-150 | **0.0182** | **0.0451** | **0.0219** |
| 1.3 | 100-100-100 | [4.5, 6.0] | 150-150-150 | 0.0218 | 0.0531 | 0.0309 |
| Cutoff radii | | | | | | |
| 1.4 | 25-50-100 | [2.0, 6.0] | 200-200-200 | 0.0347 | 0.0498 | 0.0423 |
| 1.5 | 25-50-100 | [4.5, 6.0] | 200-200-200 | 0.0465 | 0.0592 | 0.0726 |
| 1.6 | 25-50-100 | [5.0, 5.2] | 200-200-200 | 0.0408 | 0.0453 | 0.0394 |
| 1.7 | 25-50-100 | **[0.5, 6.0]** | 200-200-200 | **0.0291** | **0.0453** | **0.0383** |

| Fitting net | | | | | | |
|---|---|---|---|---|---|---|
| 1.8 | 25-50-100 | [0.5, 6.0] | 240-200-150 | 0.00664 | 0.0364 | 0.0254 |
| 1.9 | 25-50-100 | [0.5, 6.0] | 240-240-240 | 0.00203 | 0.0863 | 0.0176 |
| 1.10 | 25-50-100 | [0.5, 6.0] | **120-120-120** | **0.000264** | **0.0214** | **0.00400** |

## 3.2 Static Material Property Prediction

Since shock loading environments are the chosen application for this study, we evaluate the performance of our $B_4C$ MLIP by predicting elastic constants for small deformations and the equation-of-state (EOS) for large deformations. Elastic constants are not explicitly included in model training; thus, they can be a good preliminary indicator for model generalizability. These results are then compared to existing ReaxFF and DFT calculated values.

Using the best-performing model from Stage 1 (Model 1.10), we predict the elastic constants and EOS for the four most-prevalent $B_4C$ polytypes, with the results for the most prevalent – $B_{11}C_p(CBC)$ – and compared with available literature in Table 2 and Figure 3. Variations in elastic constant predictions can occur with differing energy cutoffs in DFT theory. Thus, we choose to use the DFT elastic constants reported in Ref. 18 as a reference point for comparison of ReaxFF and MLIP elastic performance (Table 2). This will avoid the introduction of artificial errors in ReaxFF elastic constant predictions as they will be compared to results generated with identical DFT theory. The elastic constants predicted with Model 1.10 rival the accuracy of the available ReaxFF potential in C11 and C33, while significantly improving the prediction of C12, C13, and C44. This improvement in elastic constant prediction is further supported by the accuracy of Model 1.10 in capturing the EOS in small deformation (Figure 3). Note that the current ReaxFF potential for $B_4C$ predicts artificially low atomic energies with curvature that deviates from DFT calculations in the small strain range. However, MLIP prediction of the $B_4C$ EOS diverge in highly compressive regions (< –10% strain); it is likely that the small ranges of strain (-1% to 1%) in the initial training database are beneficial to the elastic constant prediction. However, this narrow strain range proves insufficient for the prediction of the EOS in regions of large deformation.

Table 2: Comparison of elastic constants predicted by MLIP 1.10 and ReaxFF compared to DFT values from Ref [18]. Percent errors with respect to DFT values are indicated in parentheses, and bolded font indicates the lowest errors. MLIP predictions are better on average than those of ReaxFF.

|  | DFT [18] | ReaxFF [18] | MLIP 1.10 |
|---|---|---|---|
| $C_{11}$ | 572 | **501 (-12.4)** | 491 (-14.2) |
| $C_{33}$ | 535 | **521 (-2.62)** | 507 (-5.23) |
| $C_{12}$ | 122 | 238 (95.1) | **114 (-6.56)** |
| $C_{13}$ | 73 | 213 (191.8) | **112 (53.4)** |
| $C_{44}$ | 171 | 139 (-18.7) | **179 (4.68)** |

Table 2: Comparison of elastic constants predicted by MLIP 1.10 and ReaxFF compared to DFT values from Ref [18]. Percent errors with respect to DFT values are indicated in parentheses, and bolded font indicates the lowest errors. MLIP predictions are better on average than those of ReaxFF.

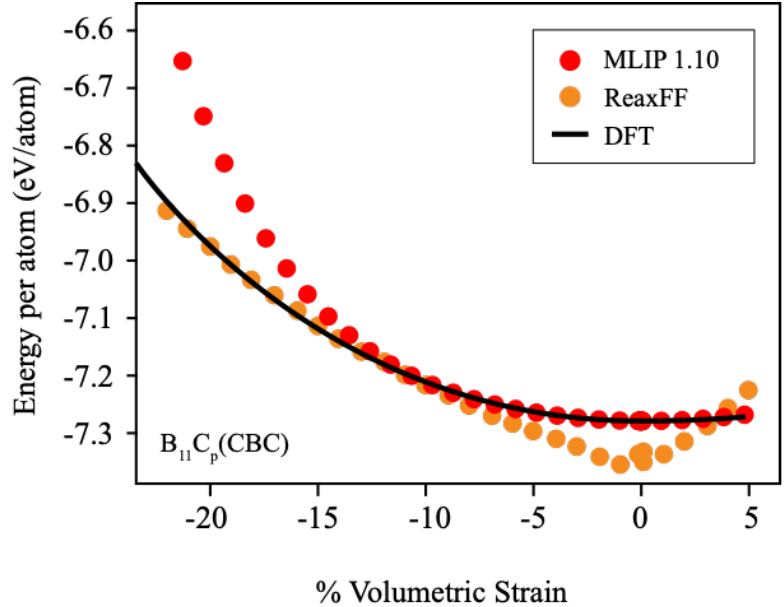

Figure 3. Comparison of energy-volume equation-of-state predicted from DFT for the most prevalent $B_4C$ polytype $B_{11}C_p$(CBC) with available ReaxFF potential and a preliminary Model 1.10. Although the MLIP is superior to ReaxFF in small strain regime, the above plot shows that the training data for the MLIP is insufficient to capture large compressive strains (beyond 12%) and motivates further refinement in training data.

Table 3. Comparison of predicted elastic constants and errors (compared to DFT) for $B_{11}C_p$(CBC) from ReaxFF and MLIPs trained on various degrees of volumetric compressive strains. Percent errors are indicated in parentheses, and bolded font indicates the lowest errors. Empty cells (denoted by '__') indicate unphysical elastic constant predictions, $C > 10^5$ GPa or $C < 0$. Energy, force, and virial stress prediction RMSE's are also reported for Models 1.10 – 3.0.

|  | DFT [18] | ReaxFF [18] | Model 1.10 1% | Model 2.0 10% | Model 2.1 20% | Model 2.2 30% | Model 2.3 40% | Model 3.0 |
|---|---|---|---|---|---|---|---|---|
| $C_{11}$ | 572 | 501 (-12.4) | 491 (-14.2) | 504 (-11.8) | 491 (-14.2) | 251 (-56.0) | — | **515 (-9.96)** |
| $C_{33}$ | 535 | 521 (-2.62) | 507 (-5.23) | **526 (-1.67)** | 508 (-4.99) | 326 (-39.2) | — | 520 (-2.73) |
| $C_{12}$ | 122 | 238 (95.1) | 114 (-6.56) | 113 (-7.61) | 120 (-2.01) | 19 (-84.6) | — | **120 (-1.7)** |
| $C_{13}$ | 73 | 213 (191.8) | 112 (53.4) | 109 (48.7) | 117 (60.8) | 143 (95.6) | — | **107 (46.7)** |
| $C_{44}$ | 171 | 139 (-18.7) | **179 (4.68)** | 189 (10.2) | 187 (9.60) | 26 (-85.0) | — | 191 (11.9) |
| Energy RMSE (eV/atom) | | | 0.000264 | 0.0100 | 0.0100 | 0.5394 | 0.2482 | 0.0116 |
| Force RMSE (eV/Å) | | | 0.0214 | 0.2619 | 0.2676 | 8.3501 | 2.1531 | 0.2773 |
| Virial RMSE (eV/atom) | | | 0.00400 | 0.0228 | 0.0229 | 0.4761 | 0.2660 | 0.0243 |

Learning from this result and noting that compressive strains experienced during shock loading of $B_4C$ can reach beyond 30% [21], the training database is adjusted to include strains up to 40% compression and 5% tension (a total of 2,714 training points) to fully encapsulate expected deformation ranges. The addition of highly compressed configurations to the training database may decrease the accuracy of the predicted elastic constants. To strike a balance between large and small deformation accuracy, four models trained on varying degrees of volumetric compression (10%, 20%, 30%, 40%) are tested and presented alongside model 1.10 in Table 3 and Figure 4. Additionally, amorphous $B_4C$ structures are included in all training datasets to capture the loss of crystallinity (amorphization) commonly observed in $B_4C$ at high pressures [19, 57]. Amorphous structures are generated by heating a 3 x 3 x 3 B12(CCC) supercell to 3675 K, and then quenching at various temperatures (3000 K, 2325 K, 1650 K, 975 K, and 300 K). The supercell is held at each temperature under an NVT ensemble for 1 ps, with snapshots taken every 1 fs resulting in an additional 6,000 total structures (5,078 training samples). MLIPs trained on databases containing up to 20%, 30%, and 40% compression (Models 2.1, 2.2, and 2.3, respectively) all predict the EOS of various $B_4C$ polytypes with sufficient accuracy compared to DFT calculations within their training regime. Notably, each model follows the DFT EOS curve for compressive strains up to approximately 10% beyond what is provided in its training database (e.g., MLIPs trained on 20% strained structures can capture configurations with up to 30% compression). Amongst the MLIPs that successfully capture the $B_4C$ EOS, only the model trained on structures with up to 20% compression can capture the elastic behavior of $B_4C$ with sufficient accuracy (Table 3). Unsurprisingly, the inclusion of highly compressed structures improves the accuracy of EOS predictions in those regions; however, the addition of too much data under extreme conditions deteriorates the prediction of elastic constants (see Model 2.3). Thus, a balance must be struck between the inclusion of sufficient strain diversity while maintaining model accuracy close to the ground-state configuration. The model trained on up to 20% compression (Model 2.1) maximizes the accuracy in compressive ranges applicable in shock ($\leq 30$ GPa) while still maintaining an acceptable error in the elastic constants. Thus, this model will proceed to the final stage of validation.

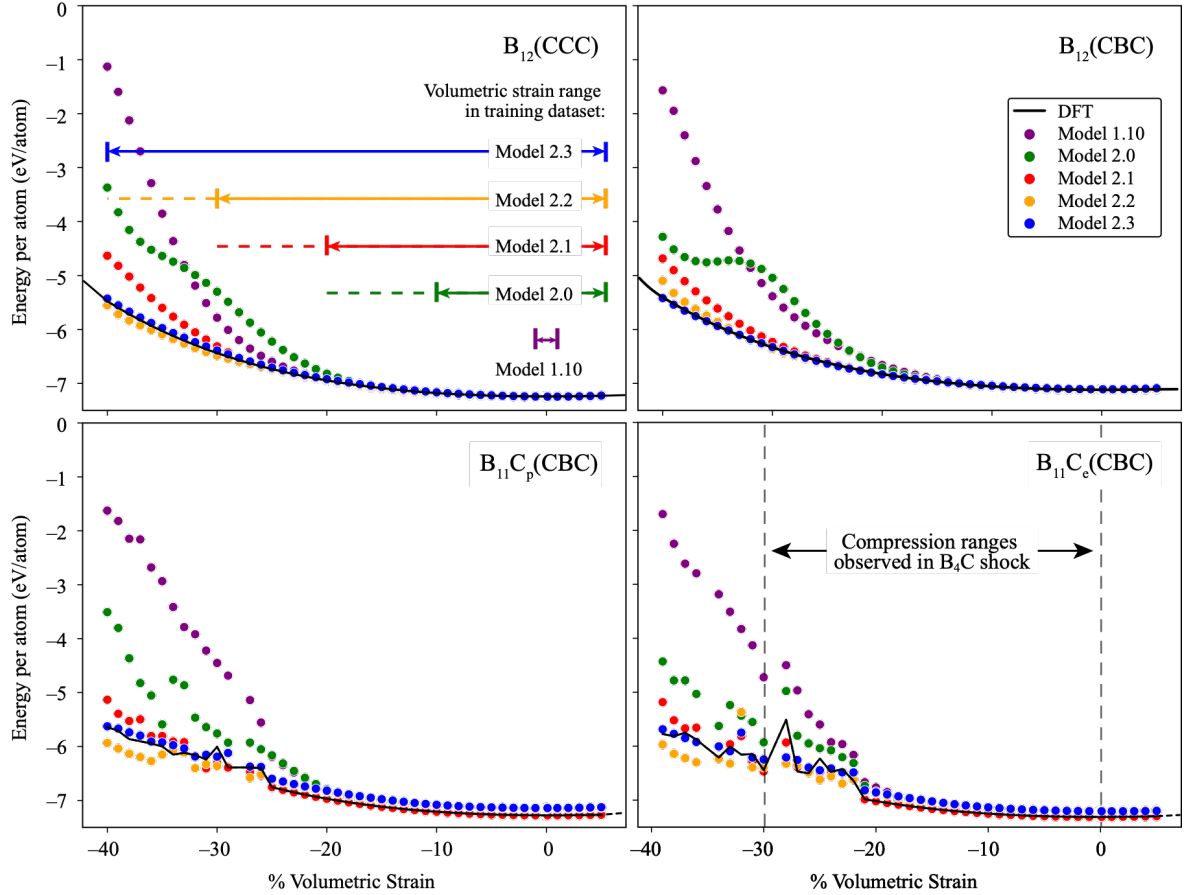

Figure 4. EOS plots for the four most prevalent $B_4C$ polytypes predicted using MLIPs trained on varying ranges of compressive volumetric strain data (10% to 40%). Results are compared to EOS predictions of DFT as well as those of model 1.10, which is trained on strains ranging from 1% compression to 1% tension. Note, the discontinuities present in the EOS of both $B_{11}C_p(CBC)$ and $B_{11}C_e(CBC)$ are speculated to be caused by structural instability in those polytypes at high pressures [14]. The horizontal solid lines in the figure for $B_{12}(CCC)$ indicate the training data range, with corresponding dashed lines showing extrapolative ability. The MLIPs can extrapolate compressive strains ~10% beyond their training range.

### 3.3 Stage 3: Dynamic property prediction

The final stage of validation aims to test MLIP performance on configurations in finite temperature environments. The dynamic movement of atoms in high-temperature environments can provide a diverse set of atomic configurations far from their ground state positions.

#### 3.3.1 Melting temperature

Due to the localized melting observed in $B_4C$ shock deformation [20, 21, 57], it is crucial to capture solid-liquid interface stability as various melt pockets in the shock domain will be surrounded by solid, crystalline material. We test the model on these conditions by predicting the melting temperature of $B_4C$ (2573 K to 2753 K [11, 58]) at ambient pressure using the interface coexistence method [59]. A fully periodic domain is split into solid and liquid regions, the liquid region is melted at 3000 K, and then the entire system is held at an equilibrium temperature ($T_e$) using the NPT ensemble to observe crystal growth (solidification) or liquid growth (melting). The temperature at which these regions are at equilibrium is taken to be the melting point. The total density of the system (2.52 g/cm³ for solid $B_4C$ [60]) is tracked over simulation time to quantify these regions, with solidification and melting resulting in increasing and decreasing densities, respectively. Figure 5(a) shows the density progression predicted using Model 2.1 along with various domain snapshots (I–III), during a temperature hold at $T_e = 1900$ K. Despite being the best performing model from stage 2 of validation, Model 2.1 fails to capture the behavior of liquid B₄C, with B and C atoms segregating into clumps, thereby forming voids (see snapshot III) and significantly decreasing the density of the system. We hypothesize this behavior is due to the lack of highly repulsive interatomic interactions (i.e., at very small separation) in the training dataset.

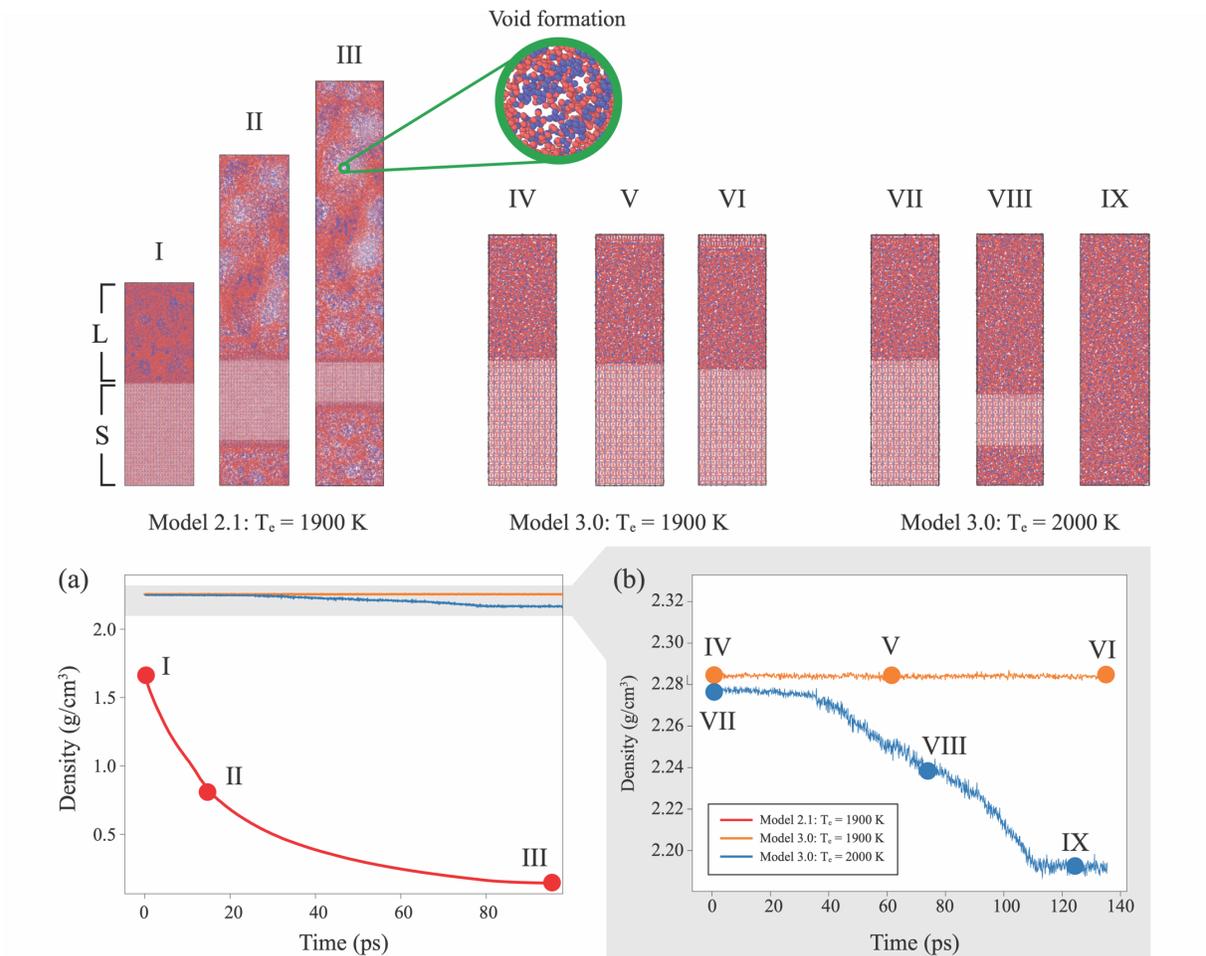

Figure 5. Domain density progressions for solid-liquid B₄C held at 1900 K and 2000 K predicted with two different models: (a) best performing model (2.1, red) from stage 2 validation predicts clumps and void formation (snapshots I–III) resulting in very low densities over time; (b) a model (3.0) trained on random atomic perturbations and α-B which reveals a stable solid-liquid interface at 1900 K (orange) and decreased density as a result of melting at 2000 K (blue). The addition of random atomic perturbation to the training dataset thus helped to capture a stable, liquid B₄C phase.

The range of interatomic distances contained in training has been implicitly determined by induced strains or thermal oscillations, neither of which would allow for separation distances small enough ($\leq 1.5$ Å) to provide sufficient examples of repulsive forces between atoms. To address this issue, random perturbations (between 0.025 Å and 0.25 Å) are performed on the atomic positions for each polytype as well as on an additional unary α-$B_{12}$ structure. Each structure is perturbed 500 times, resulting in 2500 new configurations added to the training set. This addition also increases the number of short interatomic distances (< 1.5 Å) by 27.9%. Model 3.0 is then initialized with the weights of model 2.1 and trained on this new data.

After ensuring satisfactory performance on stages 1 and 2, this updated model is used to simulate the same density progression (Figure 5(b)), at $T_e \in [1900, 2000]$ K. At $T_e = 1900$ K, Model 3.0 predicts a constant density of 2.285 g/cm³ over time of (see snapshots IV–VI), thus holding a stable solid-liquid interface. As such, the predicted melting point of single crystal $B_{12}$(CCC) using Model 3.0 is 1900 ± 50 K, which, despite being lower than reported experimental values for $B_4C$ (2573 K–2753 K [11, 58]), is within the expected error tolerance (up to 60%) seen in DFT-based melting point predictions from other ionic solids [61]. In addition to capturing a stable interface, Model 3.0 maintains a stable molten $B_4C$ phase after melting at $T_e = 2000$ K (see snapshot IX), with a liquid $B_4C$ density of 2.192 g/cm³. Despite a lack of available literature on the density of molten $B_4C$, this value is in line with expectations.

### 3.3.2 Shock Loading

Throughout the validation process, models have been tested independently on various extreme conditions (large deformations, high temperatures) that are encountered during shock loading. Hence the final step in this validation procedure is to perform shock simulations and verify that important shock physics phenomena are captured. These include capturing the pressure-volume Hugoniot and determining the Hugoniot elastic limit (HEL), which represents the transition from elastic to inelastic deformation under uniaxial strain loading. Additionally, we focus on a specific anomalous property of $B_4C$, which is its abnormal pressure-shear response beyond its HEL. For most brittle materials, their shear strength increases with pressure up to a certain limit (slightly above their HEL), after which it remains constant [62, 63]. This maximum shear strength has been shown to be proportional to the HEL for the material [64]. Based on the linear trend between the HEL and dynamic shear strength of brittle materials, the expected shear strength of $B_4C$ (HEL ≈ 18–20 GPa [65]) would be ~8.8 GPa [62, 64]. However in the case of $B_4C$, the experimental shear stress of a 98% dense sample reaches a peak (~8 GPa) near its HEL, and subsequently drops to ~3.55 GPa [66]. This loss of shear strength is due to pressure-induced amorphization during shock loading [63, 65, 67, 68]. The mechanism behind this amorphization is complex and has been linked to various factors including localized melting [20] and CBC-chain instability [16, 19, 57, 69, 70].

To determine if Model 3.0 can capture these deformation modes under the complex loading conditions present in shock environments, we employ a method similar to that in references [20, 21, 71] for shock simulations of $B_4C$. A 50.6 x 52.6 x 177.4 Å rectangular domain (64,800 atoms) with periodic lateral boundaries, is launched towards a momentum mirror with varying impact velocities, $v_z$ (see Figure 6). High resolution analysis of field quantities (pressure, volume strain, and axial strain) within the shock front is conducted using Lagrangian binning of the simulation domain along the impact direction. To avoid end-effects, a bin sufficiently far (~65.1 Å) from the impact end of the simulation cell is used for the analysis (Figure 6).

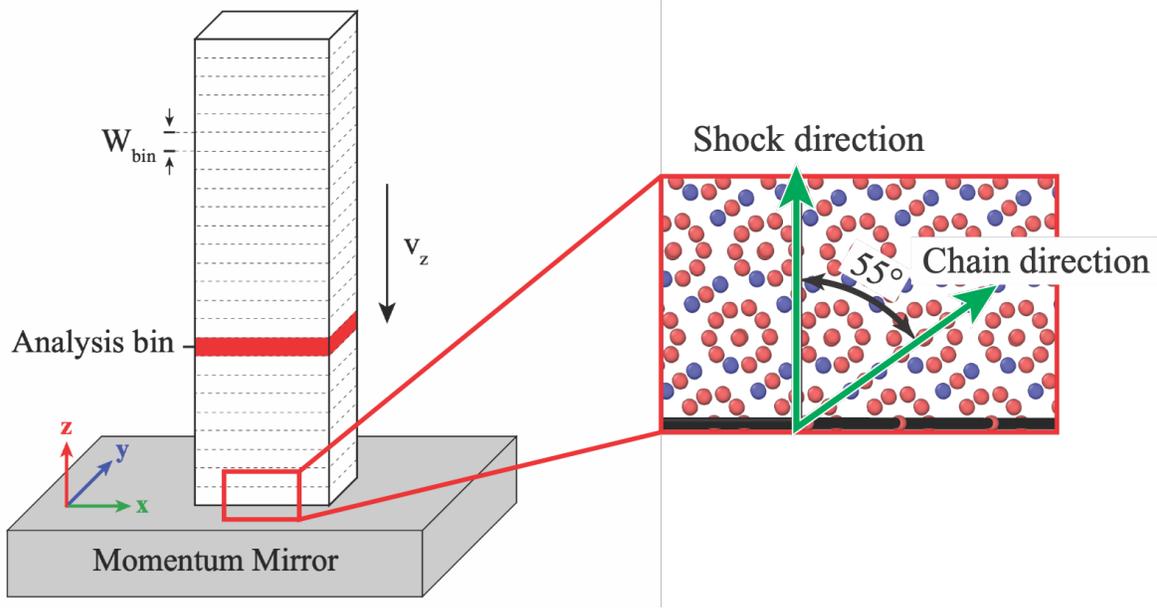

Figure 6. Lagrangian binning ($W_{bin} \approx 7.0$Å) of shock simulation cell, with a chosen analysis bin to avoid end effects while capturing shock wave in its entirety. Chain orientation is chosen at 55º to minimize the size of the cuboidal repeat unit for the $B_4C$ crystal structure.

Multiple simulations are conducted with impact velocities ranging from 0.1 km/s to 5.5 km/s using Model 3.0. The P-V Hugoniot is constructed using a similar procedure as in DeVries et al. [20], and the discontinuity in the curve is indicative of the location of the HEL. Figure 7 shows MLIP-predicted results alongside those of ReaxFF-based simulations [20], DFT calculations [70], and various experiments [72-74]. The MLIP results agree well with DFT calculations for defect-free $B_4C$ up to a pressure of 67 GPa (Fig. 7(a)), after which MLIP predictions are in closer agreement with the ReaxFF prediction by DeVries et al. [20]. However, the HEL predicted by the MLIP is ~ 67 GPa while the ReaxFF predicts the HEL between 24-60

GPa, depending on the orientation of the crystal with respect to the shock direction [21]. The ReaxFF predictions on single crystals align better with experimental values for polycrystalline $B_4C$ [65, 73] as well as DFT results for an imperfect $B_4C$ single crystal with a chain vacancy [70]. However, one would expect a higher HEL for a defect-free monocrystalline material, and the DFT calculations for defect-free $B_4C$ show a monotonically increasing trend up to 80 GPa with no visible transition from elastic to inelastic deformation. Our MLIP-based shock simulations follow this same trend, as revealed in the magnified view shown in Figure 7(b) and yield an HEL of 67 GPa. These results reveal two interesting conclusions: (i) while the ReaxFF calculations on single crystal defect-free $B_4C$ material by Devries et al., match the behavior of an imperfect material as reflected by the experimental results on polycrystalline materials, it indicates that ReaxFF-based simulations under-predict the strength in single-crystal, defect-free $B_4C$, and (ii) MLIP-based shock simulations can capture the DFT trends and yield a higher HEL value as expected for defect-free $B_4C$. Hence, the MLIP is better at capturing the expected shock physics of single-crystal $B_4C$ as compared to the existing empirical potentials.

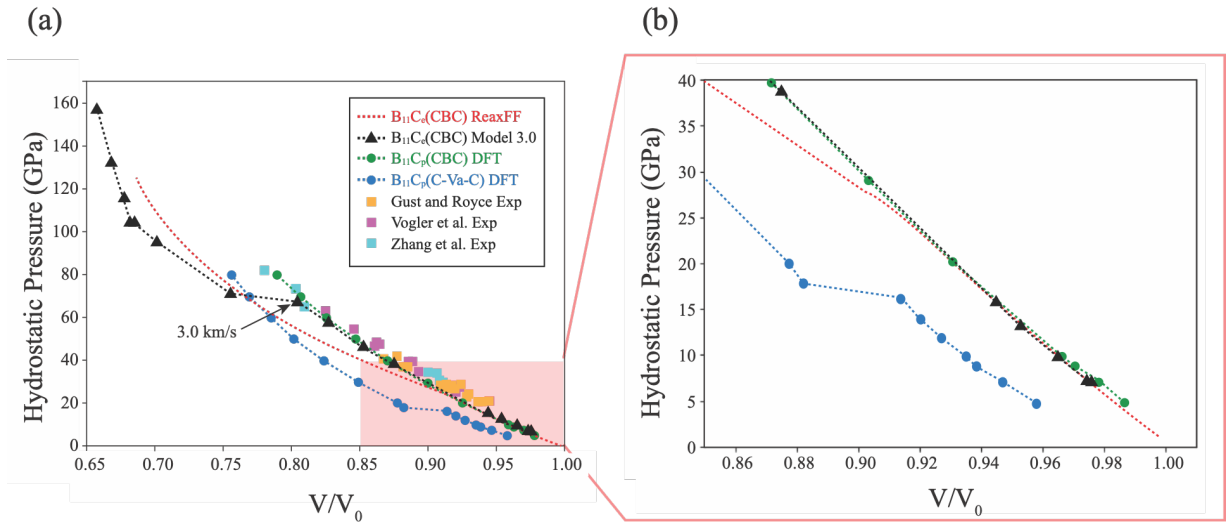

Figure 7. Pressure-volume Hugoniot relationship predicted with Model 3.0 and its comparison to DFT [70], ReaxFF [18], and experimental (Exp) results from Gust and Royce [72], Zhang et al. [74], and Vogler et al. [73]. Note, a structure with a chain vacancy ($B_{11}C_P$(C-Va-C)) is included from Taylor et al., which shows significantly lower HEL value on par with the experimental results on polycrystalline $B_4C$.

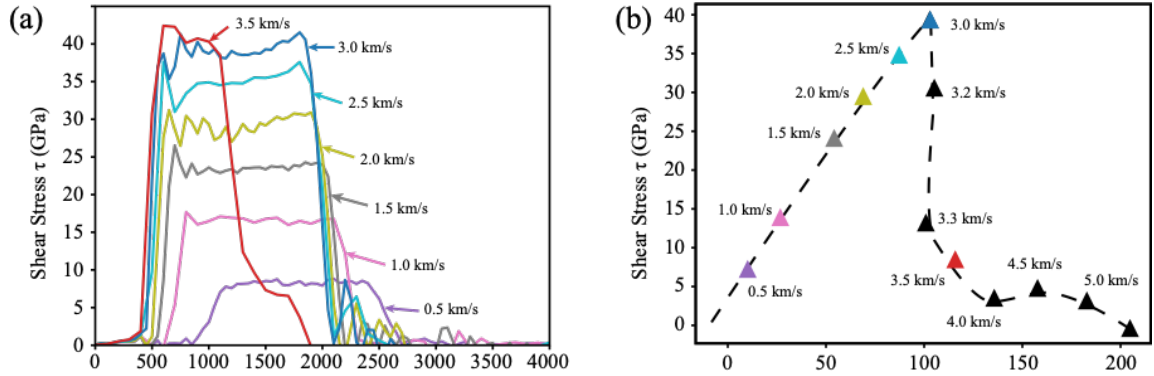

Figure 8. Shear stress, τ, within the analysis bin, as a function of (a) simulation time, and (b) axial stress in the impact direction, $\sigma_z$, for various impact velocities. Plot (b) reveals the final shear stress value at the end of the shock wave duration, with points colored according to their corresponding impact velocities as seen in plot (a).

In addition to capturing the HEL and the Hugoniot response, the MLIP can also capture the loss of shear strength during shock loading. This is demonstrated in Figure 8(a) by observing the shear stress as a function of time in the analysis bin. Note that, up to an impact velocity of 3.0 km/s, the shear stress remains constant throughout the shock duration, reflecting the ability of the shock loaded material to withstand the applied shock. However, beyond this impact velocity, the duration of the stress plateau begins to shorten, and the shear stress drops to a much lower value. The shear stress at the end of the shock duration is plotted as a function of axial stress in Figure 8(b). As expected, the maximum axial stress continuously increases with impact velocity, however the shear stress peaks at 40 GPa, followed by a steep drop-off. This behavior is similar to that observed both in experiments [62, 73] and in DFT calculations [70]. In experiments, the shear stress peaks at ~8 GPa for a polycrystalline material (with defects), while DFT calculations for a single crystal predict a shear peak at 40–50 GPa [70] which is in very good agreement with the presented MLIP results.

Our final MLIP (Model 3.0) has demonstrated accuracy across various dynamic environments by predicting the $B_4C$ melting point within a reasonable tolerance of experimental values, as well as capturing fundamental shock mechanics and loss of shear strength unique to $B_4C$. Notably, once it is able to capture melting, the MLIP did not need further adjustment to capture shock mechanics, an environment it has not seen before during the training phase. This success indicates that models trained on independent, representative subsections of the simulation environment (high temperature, large deformations, etc.) can be expected to perform well in environments where these subsections are coupled (shock loading). Thus, the transferability of the $B_4C$ MLIP is demonstrated.

## 4. Discussion

The proposed 3-stage MLIP refinement process defined in Section 3 is proven to result in an MLIP that is transferable to its desired simulation environment, while minimizing the training database size and thus reducing the computational cost of model development. The final $B_4C$ MLIP is trained on simple, representative sub-conditions that come together to form the complex loading condition of shock simulations. The success of this MLIP in shock conditions attests to its interpolative capability for conditions used in the training database. Thus, the first step of effective MLIP-development should be the breaking down of the intended simulation conditions into representative components. For example, if a user is aiming to model the diffusion kinetics of vacancies in high-pressure environments, the relevant components may be vacancy-induced structures and defect-free structures placed under large volumetric compressions and high temperatures. This can aid the user in dataset curation and refinement for their desired application. In addition to stable configurations close to equilibrium, high-energy configurations (e.g., random atomic perturbations) are shown to be essential during training to capture the full scope of the PES.

The findings in this work support the need for a robust validation procedure for ML-based interatomic potentials due to their data-driven nature. It is demonstrated that MLIPs that perform well in a numerical sense, have no guarantee of capturing the physics of a material system to the developer's desired accuracy. The continued addition of diverse structural configurations into the training database throughout the validation process also supports the claim that sufficient energetic diversity in model training is necessary for transferable MLIPs. The material-agnostic and procedural nature of this proposed workflow lends itself naturally to automation, thus an MLIP-validation package built on these concepts can significantly speed up the development time for new MLIPs for novel materials. Future works can aim to use this procedure in developing force fields for other complex ceramics.

## 5. Summary and Conclusion

The stated goals of this work were to: (i) Define a material-agnostic MLIP development process and (ii) apply this process to a well-studied ceramic system to provide a framework for expansion to other complex systems. A precise, material agnostic workflow for the development is demonstrated, with specific attention placed on the MLIP validation pathway. The proposed 3-stage process allows continuous improvements to the hyper parameters from the initial recommendations by the DeePMD-kit. The MLIP is sequentially fed through (i) preliminary statistical learning evaluation, (ii) static property prediction, and (iii) dynamic property/response prediction, adjusting the model components as necessary. This procedure offers the flexibility to choose specific metrics in each stage as well as the accuracy tolerances based on the system of interest and its desired final simulation environment, thus achieving the first goal of this work.

To meet the final objective, the viability of this process is demonstrated in its application to $B_4C$ under shock compression, showing improved alignment with DFT results for single crystals in its P-V Hugoniot as compared to previous ReaxFF studies. Further, the loss of shear strength is observed in MLIP-based MD shock simulations beyond a shear stress of 40 GPa, which agrees with HEL shear stresses of 40-50 GPa observed by Taylor et al. [70] for single crystals. Due to the GPU-optimized nature of DeePMD-kit models, this MLIP resulted in an 8-10x speedup as compared to simulations conducted with the available ReaxFF potential. Thus, a model developed using the proposed procedure is shown to capture structural and mechanical behaviors of $B_4C$, achieving the final goal of robust MLIP development for a complex structural ceramic.

## Acknowledgements

The authors acknowledge University of Florida Research Computing for providing computational resources and support that have contributed to the research results reported in this publication. URL: http://www.rc.ufl.edu.

## Data Availability

The data required to create the models, as well as the LAMMPS-compatible model files, can be found at **https://github.com/SubhashUFlorida/B4C-MLIP.git**